\documentclass[prc,twocolumn,superscriptaddress,showpacs,twoside,floatfix]%
{revtex4-1}

\usepackage{amssymb,epsfig}

\begin{document}

\title{Continuum excitations of $^{26}$O in a three-body model: $0^+$ and $2^+$
states}

\author{L.V.~Grigorenko}
\affiliation{Flerov Laboratory of Nuclear Reactions, JINR, Dubna, RU-141980
Russia}
\affiliation{National Research Nuclear University ``MEPhI'', Kashirskoye shosse
31, RU-115409 Moscow, Russia}
\affiliation{National Research Centre ``Kurchatov Institute'', Kurchatov sq.\ 1,
RU-123182 Moscow, Russia}
\author{M.V.~Zhukov}
\affiliation{Fundamental Physics, Chalmers University of Technology, S-41296
G\"{o}teborg, Sweden}


\begin{abstract}
The structure and decay dynamics for $0^+$ and $2^+$ continuum excitations of
$^{26}$O are investigated in a three-body $^{24}$O+$n$+$n$ model. The validity
of a simple approximation for the cross section profile for long-lived $2n$
emission is demonstrated. A sequence of three $0^+$ monopole (``breathing mode''
type) excited states is predicted. These states could probably be interpreted as
analogues of Efimov states pushed in the continuum due to insufficient binding.
The calculated energies of the $2^+$ states are related to the excitation
spectrum of $^{25}$O. We discuss the correlation between the predicted $^{26}$O
spectrum and experimental observations.
\end{abstract}

\pacs{ 21.60.Gx, 21.10.Tg, 21.45.-v, 24.50.+g}

\maketitle


\section{Introduction}


The interest in extremely heavy oxygen isotopes is very high today Ref.\
\cite{Hoffman:2008,Lunderberg:2012,Kohley:2013,Caesar:2013,Kondo:2014}. One
of the strongest motivations emanates from the structure theory
\cite{Volya:2006,Hagen:2009,Otsuka:2010,Hergert:2013}. The evolution of binding
energies along the oxygen isotopic chain appears to be highly sensitive to the
details of interactions and procedures used in the modern structure
approaches thus providing stringent test for their quality. This
field of research concentrates on the short distances and short-range
correlations. Recently it has been demonstrated that another source of an
inspiration here can be connected to the continuum properties of these systems
and, correspondingly, to the long-range correlations Ref.\
\cite{Volya:2006,Grigorenko:2011,Grigorenko:2013,Hagino:2014,Hagino:2014b}.

\begin{figure}
\begin{center}
\includegraphics[width=0.45\textwidth]{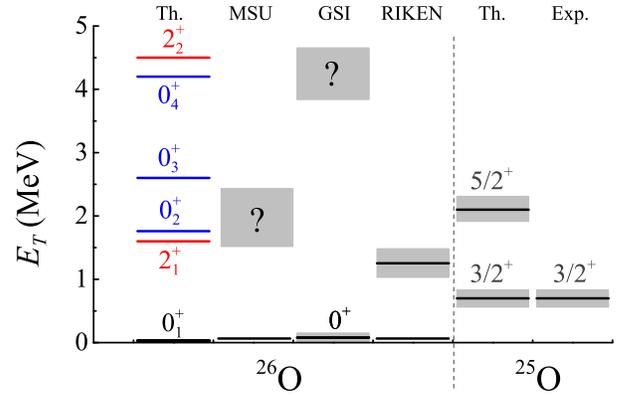}
\end{center}
\caption{(Color online) Experimentally observed energy levels of $^{26}$O
\cite{Kohley:2013,Caesar:2013,Kondo:2014} and $^{25}$O. For theory results the
level scheme predicted for $^{26}$O is shown which is based on the one assumed
for $^{25}$O.}
\label{fig:levels}
\end{figure}

The question of prospects to search for neutron radioactivity has been
considered in Ref.\ \cite{Grigorenko:2011}. In contrast to at the proton
dripline, characterized by large Coulomb barriers, in the vicinity of the
neutron dripline the emission of neutrons from the particle-unstable ground
state (g.s.) can be hindered mainly by centrifugal barriers (we do not consider
the possible structural
hindrance factors, connected with many-body effects for the ground state neutron
emitters). For isotopes with odd number of neutrons such barriers appear to be
sufficient to produce long-lived (radioactivity lifetime scale) ground states
only beyond the $s$-$d$ shell. Discussion of such a remote region of the
dripline is not of a practical importance nowadays. However, a more complicated
phenomenon comes into play for the particle-unstable dripline systems with even
number of neutrons. The pairing interaction can produce specific decay energy
conditions which force a simultaneous emission of two (or even four) neutrons.
The level schemes for $^{26}$O and $^{25}$O illustrating this situation are
given in Fig.\ \ref{fig:levels}. These, so-called true three-body (five-body)
decays, are affected by the hindrance factor connected with an appearance of
specific additional barriers in the few-body dynamics [see the discussion around
Eq.\ (\ref{eq:shredl}) below]. Thus the search for novel types of a
radioactivity phenomena, namely, two- and four-neutron radioactivities, becomes
prospective.

Among the realistic candidates for $2n$ and $4n$ radioactivity search $^{26}$O
and $^{28}$O were considered in Ref.\ \cite{Grigorenko:2011}. Since that time a
hint for the very long lifetime ($T_{1/2}=4.5^{+1.1}_{-1.5}$ ps) was obtained
for $^{26}$O in Ref.\ \cite{Kohley:2013}. Improved theoretical lifetime
estimates in Ref.\ \cite{Grigorenko:2013} indicated that an extremely low
three-body (total) decay energy, $E_T< 1$ keV, is required to produce such a
long lifetime. A realization of such a small decay energy in the nature seems
unrealistic, however, the experimental decay energy $E_T$ limit for $^{26}$O
g.s.\ was steadily decreasing towards zero in recent years. At MSU the value
$E_T=150^{+50}_{-150}$ keV was obtained \cite{Lunderberg:2012}. At GSI
$E_T<120$ keV or $E_T<40$ keV was found depending on confidence level $95 \%$ or
$68\%$, which was chosen for the analysis \cite{Caesar:2013}. In RIKEN, $^{26}$O
and $^{25}$O were also produced from $^{27}$F and $^{26}$F beams
\cite{Kondo:2014}, with much higher statistics than in Refs.\
\cite{Lunderberg:2012,Kohley:2013,Caesar:2013}. Their preliminary data also
showed the ground state just above the $^{24}$O+$2n$ threshold, in addition to
the newly observed excited state at little over 1 MeV. It should be noted that
very small decay energy ($E_T \sim 20$ keV) was predicted theoretically in Ref.\
\cite{Volya:2006}. Thus, really extreme low decay energy of this nucleus cannot
be excluded.

In recent years there were important advances in the studies of the lightest
$2p$ emitter $^{6}$Be
\cite{Grigorenko:2009,Grigorenko:2009c,Egorova:2012,Grigorenko:2012,%
Fomichev:2012}. Considering also the long history of studies of the $2n$ halo
system $^{6}$He in a three-cluster $\alpha$+$n$+$n$  approximation
\cite{Danilin:1991,Bang:1996,Danilin:2007,Ershov:2008,Ershov:2009}, it can be
concluded that the understanding of the three-body dynamics in $p$-shell nuclei
is reasonably developed. In contrast, for $s$-$d$ shell systems the situation is
much less advanced. Only very recently interesting results were obtained for the
two-proton emitter $^{16}$Ne \cite{Brown:2014,Grigorenko:2015}.
The continuum three-body dynamics of the neuron-rich $s$-$d$ shell nuclei also
remains poorly investigated with just few examples of such studies
\cite{Volya:2006,Grigorenko:2013,Hagino:2014,Hagino:2014b}, elaborating mainly
the ground state properties.

In previous works on $^{26}$O \cite{Grigorenko:2011,Grigorenko:2013} we
concentrated on the decay studies of the g.s.\ in quite schematic approaches
aiming first of all at qualitative understanding of the underlying physics for
long-living neutron emitters. In this work we consider the population and decays
of the  $^{26}$O $0^+$ and $2^+$ continuum states using more realistic model
assumptions, in a broad energy range, and with more details provided. In this
work we report several nontrivial results concerning the three-body continuum
dynamics of the $s$-$d$ shell nuclei by example of $^{26}$O, which provides us
important general insights in this question. Among the obtained results are
(i) validity of the simple approximation to the spectrum shape in the case of
radioactive $2n$ decay, (ii) existence of extreme peripheral $0^+$ monopole
excitations in the low-lying spectrum of $^{26}$O.


\section{Theoretical model}


The model applied to $^{26}$O in this work generally follows the approach of
Ref.\ \cite{Grigorenko:2013}. To provide reasonable predictions concerning the
excitation spectrum of $^{26}$O the improvements concerning the reaction
mechanism treatment were implemented, see e.g.\
\cite{Egorova:2012,Grigorenko:2015}. For some direct reactions the problem of
population and decay of three-body states can be formulated in terms of the
three-body inhomogeneous Schr\"odingier equation
\begin{eqnarray}
(\hat{H}_3 - E_T)\Psi^{(+)}_{E_T} = \Phi_{\mathbf{q}} \,,  \nonumber \\
\hat{H}_3 = \hat{T}_3 + V_{n_1\text{-}n_2} + V_{\text{core-}n_1} +
V_{\text{core-}n_2} +  V_{3}(\rho)\,, \nonumber
\label{eq:shred}
\end{eqnarray}
with the source function $\Phi_{\mathbf{q}}$ depending only on one parameter
connected to the reaction mechanism: the transferred momentum $\mathbf{q}$.
The dynamics of the three-body $^{24}$O+$n$+$n$ continuum of $^{26}$O is
described by the wave function (WF) $\Psi^{(+)}$ with pure outgoing
asymptotic in the hyperspherical harmonics (HH) formalism:
\begin{eqnarray}
\Psi^{JM_J(+)}_{E_T} = \rho^{-5/2} \sum_{K\gamma}
\chi^{(+)}_{K\gamma}(\rho)\mathcal{J}^{JM_J}_{K\gamma}(\Omega_{\rho})\,,
\nonumber
\\
\chi^{(+)}_{K\gamma}(\rho) \stackrel{\rho \rightarrow \infty}{=}
\mathcal{H}^{(+)}_{K+3/2}(\varkappa \rho) \sim \exp(+i \varkappa \rho) \, ,
\end{eqnarray}
where $\mathcal{H}$ denote the Riccati-Bessel functions of half-integer index
and the ``multi-index'' $\gamma$  denotes the complete set of quantum
numbers except for the principal quantum number $K$: $\gamma=\{L,S,l_x,l_y\}$.

The three-body calculations in the HH method utilize the transition from the
three-body Jakobi coordinates
$\{\mathbf{x},\mathbf{y}\}=\{x,\Omega_x,y,\Omega_y\}$ to the collective
coordinates $\{\rho,\Omega_{\rho}\}=\{\rho,\theta_{\rho},\Omega_x,\Omega_y\}$.
The hyperradius $\rho$ (describing collective radial motion) and
the hyperangle $\theta_{\rho}$ (responsible for geometry of the system at given
$\rho$) are defined via the cluster coordinates $\mathbf{r}_i$ as:
\begin{eqnarray}
\mathbf{x} &=& \sqrt{\textstyle \frac{A_1A_2}{A_1+A_2}} (\mathbf{r}_1 -
\mathbf{r}_2) , \nonumber  \\
\mathbf{y}  &=& \sqrt{\textstyle \frac{(A_1+A_2)A_3}{A_1+A_2+A_3}} \left(
\textstyle \frac{A_1\mathbf{r}_1+A_2\mathbf{r}_2}{A_1+A_2} - \mathbf{r}_3
\right),
\nonumber \\
\rho &=& \sqrt{x^2+y^2} \;, \quad \theta_{\rho}=\arctan(x/y) \, .
\label{eq:coord}
\end{eqnarray}
Hypermomentum $\varkappa=\sqrt{2ME_{T}}$ is the dynamic variable conjugated to
hyperradius. The mass $M$ is an average nucleon mass for the considered nucleus.

The hyperspherical harmonics $\mathcal{J}^{JM_J}_{K\gamma}$ with definite total
angular momentum $J$ and its projection $M_J$
\[
\mathcal{J}^{JM_J}_{K\gamma}(\Omega_{\rho}) =
\psi^{l_xl_y}_L(\theta_{\rho})\,\left[ [Y_{l_x}(\Omega_x) \otimes
Y_{l_y}(\Omega_y)]_L
\otimes X_S \right]_{JM_J}
\]
form a full set of orthogonal functions on the five-dimensional ``hypersphere''
$\Omega_{\rho}$. The pure hyperangular functions $\psi^{l_xly}_L$ are expressed
in terms of Jacobi polinomials.

The three-body Schr\"odinger equation (\ref{eq:shred}) in the hyperspherical
basis is reduced to the set of coupled differential equations for the functions
$\chi ^{(+)}$:
\begin{eqnarray}
\left[ \frac{d^{2}}{d\rho ^{2}} - \frac{\mathcal{L}(\mathcal{L}+1)}{\rho ^{2}}
+2M\left\{ E-V_{K\gamma ,K\gamma }(\rho )\right\} \right] \chi _{K\gamma
}^{(+)}(\rho ) \nonumber \\
=2M \!\!\! \sum_{K^{\prime }\gamma ^{\prime } \neq K \gamma }\!\! V_{K\gamma,
K^{\prime } \gamma ^{\prime
}}(\rho )\chi _{K^{\prime }\gamma ^{\prime }}^{(+)}(\rho ) - 2M\,\Phi
_{\mathbf{q},K\gamma }(\rho )\,, \quad
\label{eq:shredl}
\end{eqnarray}
which can be interpreted as motion of a single ``effective'' particle in a
strongly deformed field. The ``three-body potentials'' (matrix elements of the
pairwise potentials) $V_{K\gamma ,K^{\prime }\gamma ^{\prime}}(\rho )$ and the
partial source terms $\Phi_{\mathbf{q},K\gamma}$ are defined as
\begin{eqnarray}
V_{K\gamma ,K^{\prime }\gamma ^{\prime }}(\rho ) & = &
\int \!d\Omega _{\rho }\,\mathcal{J}_{K^{\prime }\gamma ^{\prime
}}^{JM_J\ast }(\Omega _{\rho}) \sum_{i<j}V_{ij}(\mathbf{r}_{ij})
\,\mathcal{J}_{K\gamma }^{JM_J}(\Omega _{\rho})\,, \nonumber \\
\Phi_{\mathbf{q}, K\gamma }(\rho ) & = & \int \! d \Omega _{\rho }\,
\mathcal{J}_{K^{\prime} \gamma ^{\prime}}^{JM_J \ast}(\Omega _{\rho}) \,
\Phi_{\mathbf{q}}(\rho,\Omega _{\rho})\,. \nonumber
\label{eq:hhpot}
\end{eqnarray}
The details of the hyperspherical method application to various three-body
systems in different physical situations can be found in the papers
\cite{Danilin:1991,Grigorenko:1998,Grigorenko:2001,Grigorenko:2008,%
Grigorenko:2009c}.

The important qualitative difference between Eq.\
(\ref{eq:shredl}) and the conventional two-body situation is that the
``effective angular momentum'' $\mathcal{L}=K+3/2$ in the three-body problem is
not equal to zero even for the lowest possible quantum state with $K=0$. Thus,
there exists a three-body centrifugal barrier even for decays via $s$-wave
emission of neutral particles producing strong hindrance factors for the widths
of such low-energy decays.

The differential cross section is expressed via the flux $j$ induced by the WF
$\Psi^{(+)}$ on the remote five-dimensional surface $\Omega_{\rho}$ with
$\rho=\rho_{\max}$
\begin{eqnarray}
\frac{d\sigma}{dE_T\, d\Omega_{\rho}} \sim j, \qquad j = \left. \langle
\Psi^{(+)}_{E_T} | \hat{j} | \Psi^{(+)}_{E_T} \rangle \right|_{\rho_{\max}}
\nonumber \\
= \frac{1}{M} \, \text{Im} \left. \left[ \Psi^{(+)\dagger}_{E_T} \,\rho^{5/2}
\frac{d}{d\rho}\rho^{5/2}\, \Psi^{(+)}_{E_T} \right]  \right|_{\rho_{\max}}\,.
\label{eq:cross-sect}
\end{eqnarray}

The approach with inhomogeneous Schr\"odinger equation (\ref{eq:shred}) had
previously been applied to two different direct reaction mechanisms (knockout
and charge-exchange) populating the three-body continuum of the $^{6}$Be
\cite{Egorova:2012,Grigorenko:2012,Fomichev:2012}, $^{10}$He
\cite{Sharov:2014}, and $^{16}$Ne \cite{Brown:2014,Grigorenko:2015}.

The source function $\Phi_{\mathbf{q}}$ for the $0^{+}$ continuum was
approximated assuming a sudden removal of a $d$-wave proton from $^{27}$F
\begin{equation}
\Phi_{\mathbf{q}}^{(0^+)} = v_0 \int d^3 r_p e^{i\mathbf{q r}_p} \langle
\Psi_{^{24}\text{\scriptsize O}} | \Psi_{^{27}\text{\scriptsize F}} \rangle \, ,
\label{eq:sour-0p}
\end{equation}
where $\mathbf{r}_p$ is the radius-vector of the removed proton. The $^{27}$F
g.s.\ WF $\Psi_{^{27}\text{\scriptsize F}}$ was obtained in a three-body
$^{25}$F+$n$+$n$ cluster model and the technicalities of proton removal from
the $^{25}$F core of the $^{27}$F nucleus are the same as in calculations of
Ref.\ \cite{Sharov:2014}.

The source function $\Phi_{\mathbf{q}}$ for the $2^{+}$ continuum can not be
easily evaluated by a simple proton removal model in the framework of three-body
approach to structure of $^{27}$F. Therefore, we use the source generated by
additionally acting on the valence neutrons of the $^{27}$F g.s.\ WF by the
quadrupole operator:
\begin{equation}
\Phi_{\mathbf{q}}^{(2^+)} = v_2 \int d^3 r_n e^{i\mathbf{q r}_n} \langle
\Psi_{^{24}\text{\scriptsize
O}} |\sum_{i=1,2} r^2_i \, Y_{2m_i}(\hat{r}_i)| \Psi_{^{27}\text{\scriptsize F}}
\rangle \, .
\label{eq:sour-2p}
\end{equation}
Expressions of this type typically arise in the direct reaction studies and seem
to be sufficiently suited for exploratory studies of $2^+$ excitations in the
$^{26}$O case. The approximation is the same as used in the recent  paper
\cite{Grigorenko:2015}.

The sudden removal approximation is not intended for absolute cross section
calculations, and therefore the source strength coefficients $v_i$ in Eq.\
(\ref{eq:sour-0p}) and (\ref{eq:sour-2p}) are arbitrary values providing the
source function the correct dimension of $[\text{energy}/\text{length}^{5/2}]$.


\section{Potentials}


In this work we employ for the nucleon-nucleon channel
the quasirealistic potential from Ref.\ \cite{Gogny:1970} including central,
spin-orbit, tensor, and parity-splitting terms.

In the $^{24}$O-$n$ channel we used the
potential from \cite{Grigorenko:2013} characterized as ``moderate repulsion in
$s$ and $p$ waves''. This is a Woods-Saxon potential with radius $r_0=3.5$ fm,
diffuseness $a=0.75$ fm, and depth parameters $V_s=70$ MeV and $V_p=70$ MeV for
$s$ and $p$ waves respectively. The $d$-waves component was modified by an
inclusion of the $ls$ interaction to produce realistic excitation spectrum of
$^{25}$O: $V_d=-33$ MeV, $V_{ls}=-5$ MeV. We also varied the $ls$ interaction to
check the sensitivity to this parameter, see Sections \ref{sec:monopole} and
\ref{sec:2p}.

The three-body potential $V_{3}(\rho)$ depending on $\rho$ with the Woods-Saxon
parameterization ($\rho_0=5$ fm, $a=0.9$ fm)  is used to control the decay
energy $E_T$ when the fine adjustment is needed, see the discussion in
\cite{Pfutzner:2012}. This potential is very small (on the level of just a few
keV) for the $^{24}$O-$n$ potential chosen in this work and therefore $V_{3}$
does not influence the other results of the calculations on the practical level.


\section{Excitation spectrum and structure of $0^+$ states}
%

\begin{figure}
\includegraphics[width=0.47\textwidth]{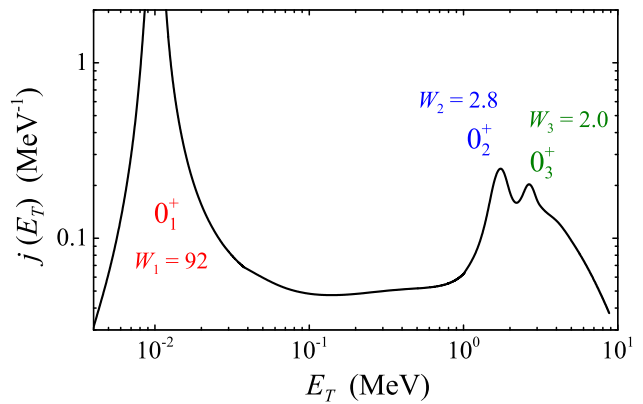}
\caption{(Color online) The $0^+$ excitation spectrum of $^{26}$O calculated by 
the three-body model for proton knockout from $^{27}$F. Weights, $W_i$, of the 
peaks are shown in percent.}
\label{fig:spec-0p}
\end{figure}

A calculated spectrum for $0^+$ continuum states of $^{26}$O is shown in Fig.\
\ref{fig:spec-0p}. In this particular calculation the g.s.\ is obtained at
$E_T=10$ keV. Two more states are obtained at about 1.76 and 2.6 MeV. Some hint
for a broad $0^+$ state at about 4.2 MeV can also be seen. Within the simple
reaction model used in this work the ground $0^+$ state is populated with the
relative intensity $W_{1} \sim 92\%$ indicated in Fig.\ \ref{fig:spec-0p}, while
the excited $0^+$ states are populated at a level of $W_{i>1}  \sim  2-3\;\%$.
It can be expected that in more complicated reaction scenario the source
function $\Phi_{\mathbf{q}}$ should have smaller affinity to the $^{26}$O ground
state WF and we can expect up to $W_{i>1} \sim 5-15 \;\%$ population rates for
the excited $0^+$ states. The relative population intensities above are
calculated by integration of the strength function Fig.\ \ref{fig:spec-0p} in
the range $[E_{3r}(i)-\Gamma(i),E_{3r}(i)+\Gamma(i)]$ around the energy
$E_{3r}(i)$ of the $i$-th resonance which width is $\Gamma(i)$.

\begin{table}[b]
\caption{The three-body structure of the g.s.\ of initial $^{27}$F nucleus and
$^{26}$O $0^+$ excitations in terms of probability $W(l^2)$ in percent of the
corresponding $[l^2_j]_0$ configuration. The energies of the states with respect
to the three-body breakup threshold are provided in the last row.}
\begin{ruledtabular}
\begin{tabular}[c]{ccccc}
$l_j$  & $^{27}$F, g.s.\  &  $^{26}$O, g.s.\ & $^{26}$O, $0^+_2$ & $^{26}$O,
$0^+_3$ \\
\hline
$s_{1/2}$   & 0.55   &  0.67 & 3.7 & 3.8 \\
$d_{3/2}$   & 84     &  79  & 80  & 86  \\
$d_{5/2}$   & 13     &  19  & 6.0   & 6.1   \\
\hline
$E_T$ (MeV) & $-3.2$ & 0.01 &  1.7 & 2.6
\end{tabular}
\end{ruledtabular}
\label{tab:0p-struct}
\end{table}


\section{R-matrix like phenomenology}


The ideas of using R-matrix type expressions for analysis of the three-body
excitations and decays have been discussed occasionally in the literature
\cite{Golovkov:2004,Grigorenko:2008,Grigorenko:2009} in quite a sketchy way. It
is important to understand general validity and limits of applicability of such
expressions for practical application as for theoretical estimates and as for
phenomenological analysis of experimental data.

It can be shown that for the source function normalized at given $\mathbf{q}$
value
\[
\int d \rho \, \rho^5 \, d\Omega_{\rho}  \,2M \, |\Phi_{\mathbf{q}}
(\rho,\Omega_{\rho}) |^2 = 1\, ,
\]
the energy profile of flux is provided by the expression
\begin{eqnarray}
j(E_T) & = & W(E_{3r}) \, \frac{\pi}{2M^2} \,
\frac{\Gamma_K(E_T)}{(E_T-E_{3r})^2 - \Gamma(E_{3r})^2/4}\,, \nonumber \\
\Gamma_K(E_T) & = & 2 \, \gamma_{\text{\scriptsize WL}}\, \theta^2_{K \gamma} \,
\frac{(2/\pi)}{J^2_{K+2}(\varkappa \rho_{\text{ch}}) + Y^2_{K+2}(\varkappa
\rho_{\text{ch}})} \,.
\label{eq:r-matr}
\end{eqnarray}
Here the hypermomentum $\varkappa=\sqrt{2ME_{3r}}$ is defined at the three-body
resonance energy $E_{3r}$, and the ``Wigner limit'' $\gamma_{\text{\scriptsize
WL}}=1/(2M\rho^2_{\text{ch}})$ estimates the upper limit for the width. The
functions $J$ and $Y$ are cylindrical Bessel functions regular and irregular at
the origin, respectively. This expression is totally analogous to the standard
two-body R-matrix expression as it is based on the assumption that the
penetration in the three-body system is defined by the three-body channel in the
hyperspherical decomposition of the WF with the lowest possible hypermomentum
$R=K_{\min}$. For systems with zero spin of the core the latter is trivially
related to the total spin of the state: $K_{\min}=J$. The ``normalization''
$W(E_{3r})$ is connected to the affinity of the source to the inner structure of
the resonance; this value is expected to be smaller than unity, but of the order
of unity for the realistic situation. It was found to be $W(\text{g.s.}) = 0.92$
in our calculations, see also Fig.\ \ref{fig:spec-0p}.

\begin{figure}
\includegraphics[width=0.238\textwidth]{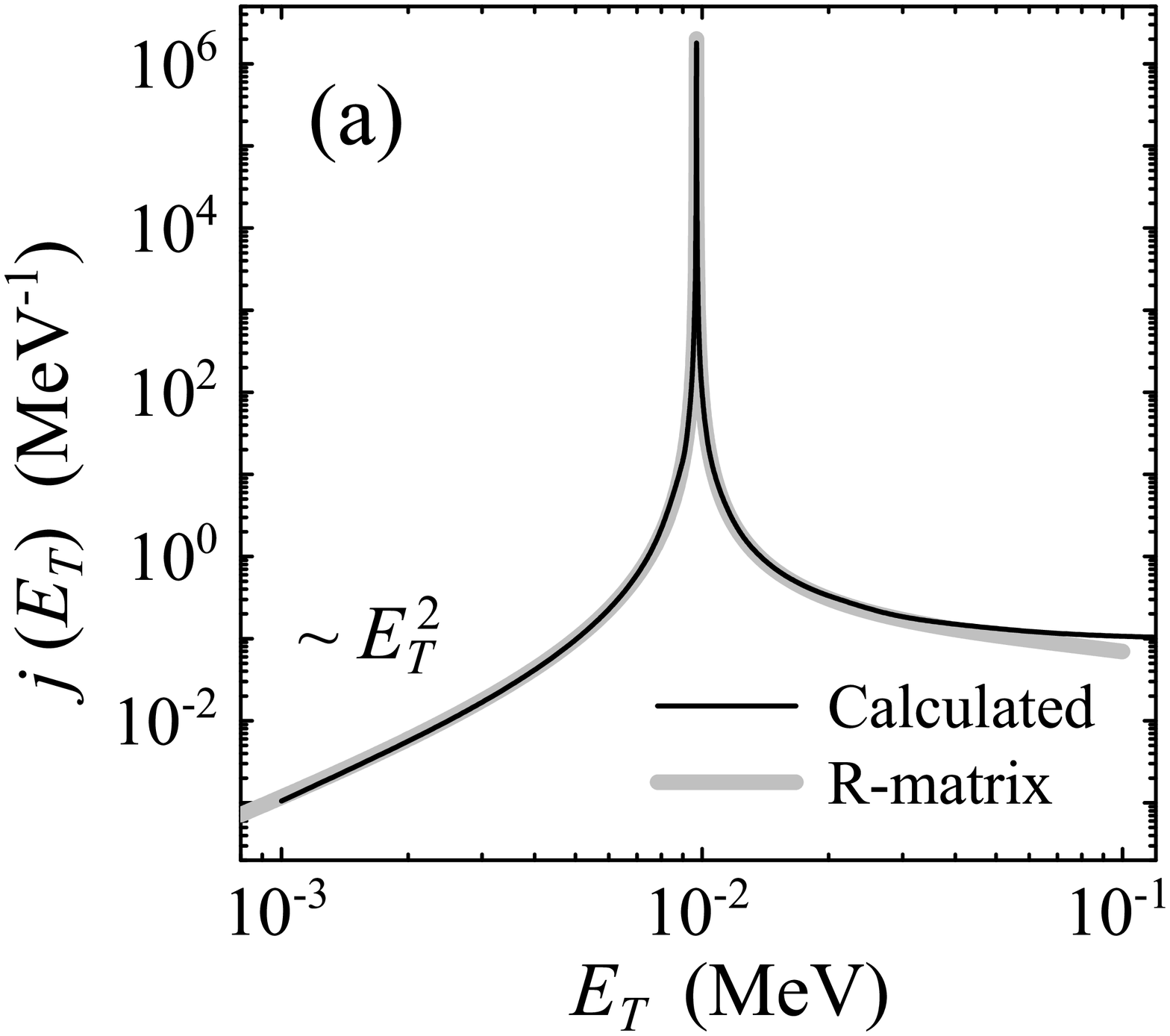}
\includegraphics[width=0.238\textwidth]{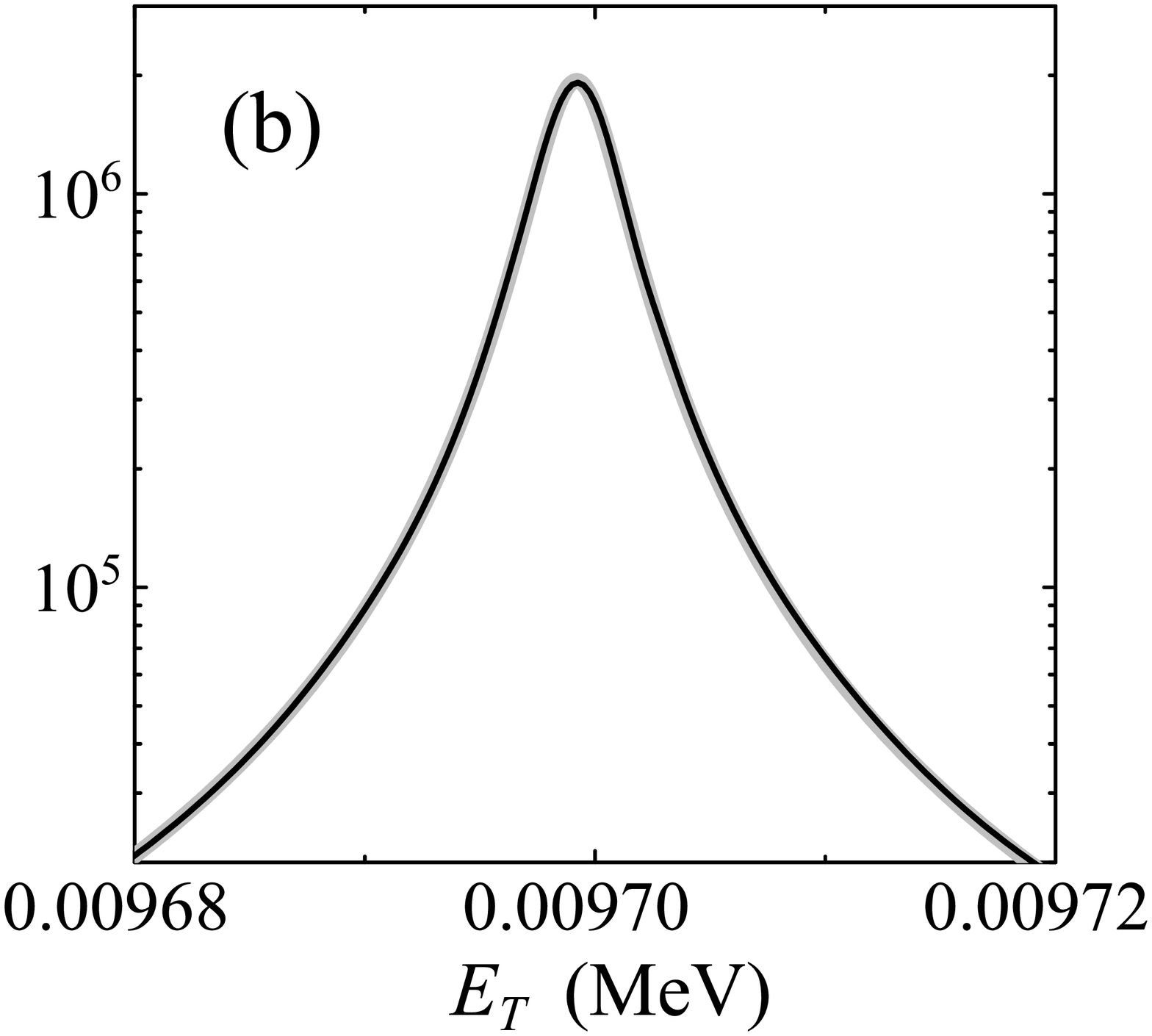}
\caption{Cross section  profiles for the very narrow $^{26}$O g.s.\ calculated
by the three-body model and obtained by the R-matrix-type approximation are
shown on two different scales in panels (a) and (b).}
\label{fig:r-matr}
\end{figure}

The example of application of the R-matrix expression is provided in Fig.\
\ref{fig:r-matr}. The possibility of fitting the excitation profile by the
Breit-Wigner shape in the vicinity of the resonance does not cause any doubts.
However, it can be seen that the near-perfect description is provided by the
R-matrix expression up to something like $10^5$ widths away from the resonance.
For this Figure the ``spectroscopic factor '' $\theta^2_{K\gamma}$ is fitted in
such a way that the calculated width $\Gamma_K$ for the three-body resonance
energy $E_{3r} = 9.7$ keV is exactly reproduced by the R-matrix expression [in
this specific case $\Gamma_K(E_{3r}) = 4.04 \times 10^{-3} $ keV].

\begin{figure}
\includegraphics[width=0.47\textwidth]{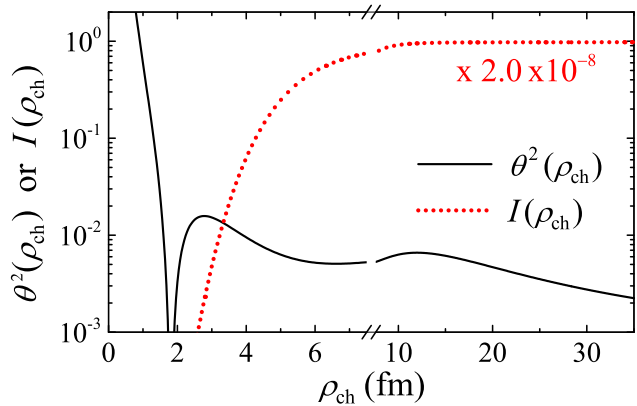}
\caption{(Color online) Spectroscopic factor  $\theta^2$ Eq.\ (\ref{eq:r-matr}) 
as a function of channel radius $\rho_{\text{ch}}$. The dotted curve shows the 
dependence of the ``internal normalization'' $I$ on $\rho_{\text{ch}}$.}
\label{fig:red-wid}
\end{figure}

It is also necessary to fix one more parameter: the ``channel radius''
$\rho_{\text{ch}}$. In the three-body case this parameter does not have such a
well defined meaning as in two-body R-matrix phenomenology and some
investigation is required here. Definition of the spectroscopic factor
$\theta^2$ for the single-channel approximation is
\[
\theta^2_{K \gamma}(\rho_{\text{ch}}) =
\frac{|\chi_{K\gamma}(\rho_{\text{ch}})|^2} {I( \rho_{\text{ch}})}\, , \;
I( \rho_{\text{ch}}) = \sum_{K \gamma} \int_0^1 \!\! dx \, |\chi_{K\gamma}( x
\rho_{\text{ch}})|^2 .
\]
The $\theta^2$ dependence on channel radius, obtained using the calculated
three-body $^{26}$O WF, is illustrated in Fig.\ \ref{fig:red-wid}. It is very
stable in a broad range of channel radii,
varying from 0.007 to 0.004 for $4<\rho_{\text{ch}}<25$ fm. The $\theta^2$
variation for the mentioned range of channel radius is in very good agreement
with the relative weight of the $[s^2]$ configuration calculated within the
whole internal region $W(s^2)=0.0067$, see also Table \ref{tab:0p-struct}. The
calculated three-body width is reproduced for
$\rho_{\text{ch}}=13$ fm and for less than $\pm 50 \%$ variation of width we
need to keep the channel radius in the range $9<\rho_{\text{ch}}<16$ fm.

We can see that the extension of a simple single channel R-matrix phenomenology,
Eq.\ (\ref{eq:r-matr}), to the three-body decays provides easily tractable and
very reliable results for long-lived two-neutron emitters. We can also conclude
that the use of typical R-matrix parameters, chosen according to the
prescription discussed above, can be expected to provide widths values with an
uncertainty around $50\%$, which is a quite accurate result for estimates
concerning long-living states.


\section{The monopole $0^+$ excitations}
\label{sec:monopole}


The nature of the excited $0^+$ states is very interesting and deserves a
special discussion. Table \ref{tab:0p-struct} provides the basic structure
information about the $^{26}$O $0^+$ states indicating their high similarity. It
is clear that a simple explanation for such a situation is that the predicted
excited $0^+$ states are all monopole (often called ``breathing mode'')
excitations.

\begin{figure}
\includegraphics[width=0.47\textwidth]{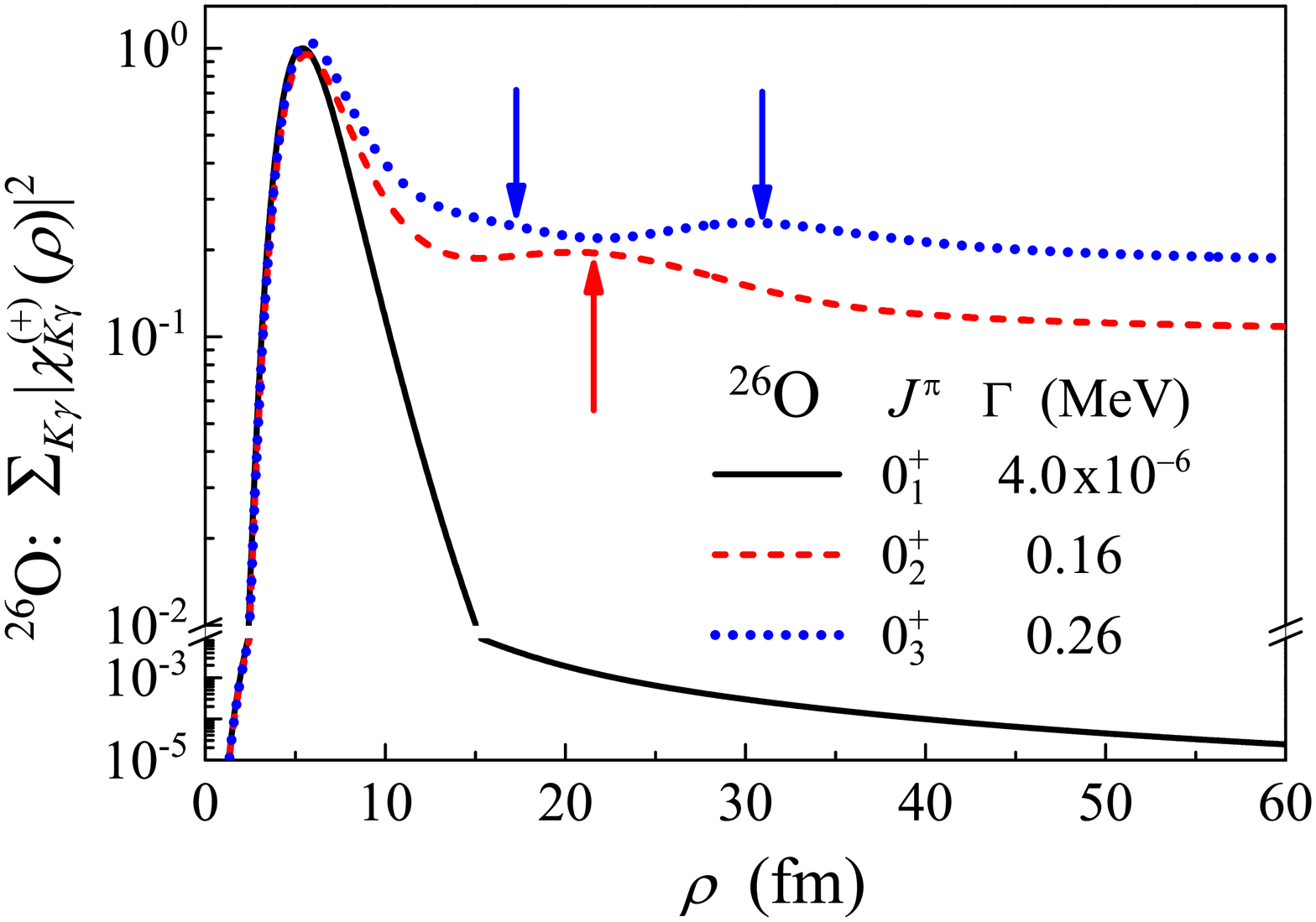}
\caption{(Color online) WF densities for the $^{26}$O ground state and two 
excited $0^+$ states. All WFs are normalized to unity maximum value. Arrows 
indicate the humps connected with monopole excitations.}
\label{fig:wf-den-mono}
\end{figure}

To check this assumption we have studied the radial evolution and the
correlation densities of the $^{26}$O $0^+$ WFs at corresponding energies. One
can see in Fig.\ \ref{fig:wf-den-mono} that the g.s.\ WF density decreases more
or less exponentially inside the barrier. At larger distances the behavior tends
to be constant which corresponds to approaching the asymptotic behavior
$\chi^{(+)} \sim \exp(+i\varkappa \rho)$ of the three-body WF. However, for the
$0^+_2$ and $0^+_3$ state WFs there exists one and two extra humps respectively
(indicated by arrows in Fig.\ \ref{fig:wf-den-mono}) in the above-the-barrier
slope of the density before the asymptotic (constant) behavior is achieved. This
is exactly expected for the monopole states where the major WF component should
have one or more nodes in the radial WF. The extreme radial extent of the humps
is notable. E.g.\ the peak for $0^+_3$ with $\rho \sim 30$ fm corresponds to
typical single particle distances in the core-neutron channel of about 20 fm.

\begin{figure}
\includegraphics[width=0.47\textwidth]{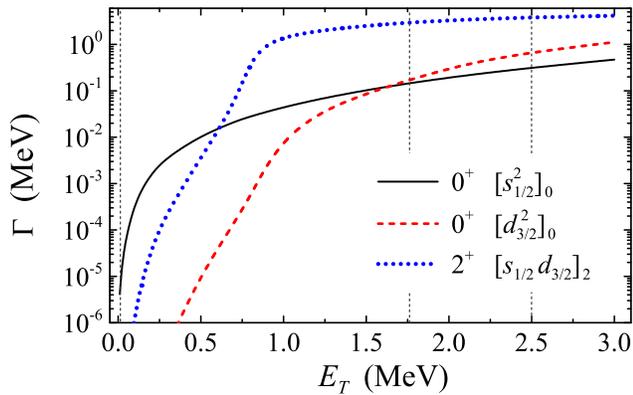}
\caption{(Color online) Width estimates for the $0^+$ and $2^+$ states of 
$^{26}$O in the R-matrix-type approach of Eq.\ (\ref{eq:sequent}) for different 
$[j_1j_2]_J$ configurations. Vertical dashed lines indicate the predicted 
positions of the $0^+$ states. }
\label{fig:wid-estim}
\end{figure}

To understand the reasons for the monopole state formation we performed simple
width estimates for major configurations of the $^{26}$O WF.
Fig.\ \ref{fig:wid-estim} provides the estimates in a ``direct decay''
R-matrix model from Ref.\ \cite{Pfutzner:2012}:
\begin{eqnarray}
\Gamma_{j_1 j_2}(E_T) =  \frac{E_{T}\left \langle V_{3}\right \rangle
^{2}} {2\pi} \int_0^1 \!\! d \varepsilon \,
 \frac{ \Gamma_{j_1}(\varepsilon E_{T})}
{(\varepsilon E_{T}-E_{j_1})^{2}+\Gamma_{j_1}(\varepsilon E_{T})^{2}/4}
\nonumber \\
 \times  \frac{\Gamma_{j_2}((1-\varepsilon)E_{T})}
{((1-\varepsilon)E_{T}-E_{j_2})^{2} + \Gamma_{j_2}
((1-\varepsilon)E_{T})^{2}/4}\;.\qquad
\label{eq:sequent}
\end{eqnarray}
This model is reasonably well approximating the true three-body decay mechanism,
and also provides a smooth transition to the sequential decay regime. The direct
decay model is constructed in the spirit of independent particle approach with
two nucleons being emitted from states with definite single-particle angular
momenta $j_1$ and $j_2$ sharing the total decay energy $E_T$, but with
interaction between nucleons neglected. The $\Gamma_{j_i}$ is the standard
R-matrix expression for the width as a function of the energy for the involved
resonances in the $^{24}$O+$n$ subsystems. It is assumed that two-body resonant
states with energies $E_{j_i}$ are present in both two-body subsystems and that
the values $\Gamma_{j_i}(E_{j_i})$ correctly describe their empirical widths.
The matrix element $\left \langle V_{3}\right \rangle$ can be well approximated
as
\[
\left \langle V_{3}\right \rangle ^{2}   =   D_3 [(E_T-E_{j_1}-E_{j_2})^2 +
(\Gamma_{j_1}(E_{j_1})+\Gamma_{j_2}(E_{j_2}))^2/4]\, ,
\]
where the parameter $D_3 \approx 1.0-1.5$ is a constant.

It can be seen in Fig.\ \ref{fig:wid-estim} that for low-energy decays of the
$0^+$
states the predominant contribution to the width is connected with the
$[s_{1/2}^2]_0$ configuration of $^{26}$O. Here the width associated with the
dominant $[d_{3/2}^2]_0$ configuration of the WF (see Table \ref{tab:0p-struct})
is strongly suppressed. However, for $E_T>0.77$ MeV the decay mechanism for the
$[d_{3/2}^2]_0$ configuration changes to the sequential emission via the
$d_{3/2}$ ground state of $^{25}$O. So, at higher energies the partial width for
this configuration rapidly grows and becomes approximately equal to that of
the $[s_{1/2}^2]_0$ configuration at $E_T=1.6$ MeV. This is approximately the
energy at which the second $0^+$ state appears. So, despite the simplicity of
these estimates they provide a hint that the appearance of the excited $0^+$
states could be connected with the ability of the $[d_{3/2}^2]_0$ configuration
to propagate effectively to large distances above the barrier. We can estimate
that the monopole states found, are built on interference patterns between
$[s_{1/2}^2]_0$ and $[d_{3/2}^2]_0$ components at large distances.

\begin{figure}
\begin{center}
\includegraphics[width=0.49\textwidth]{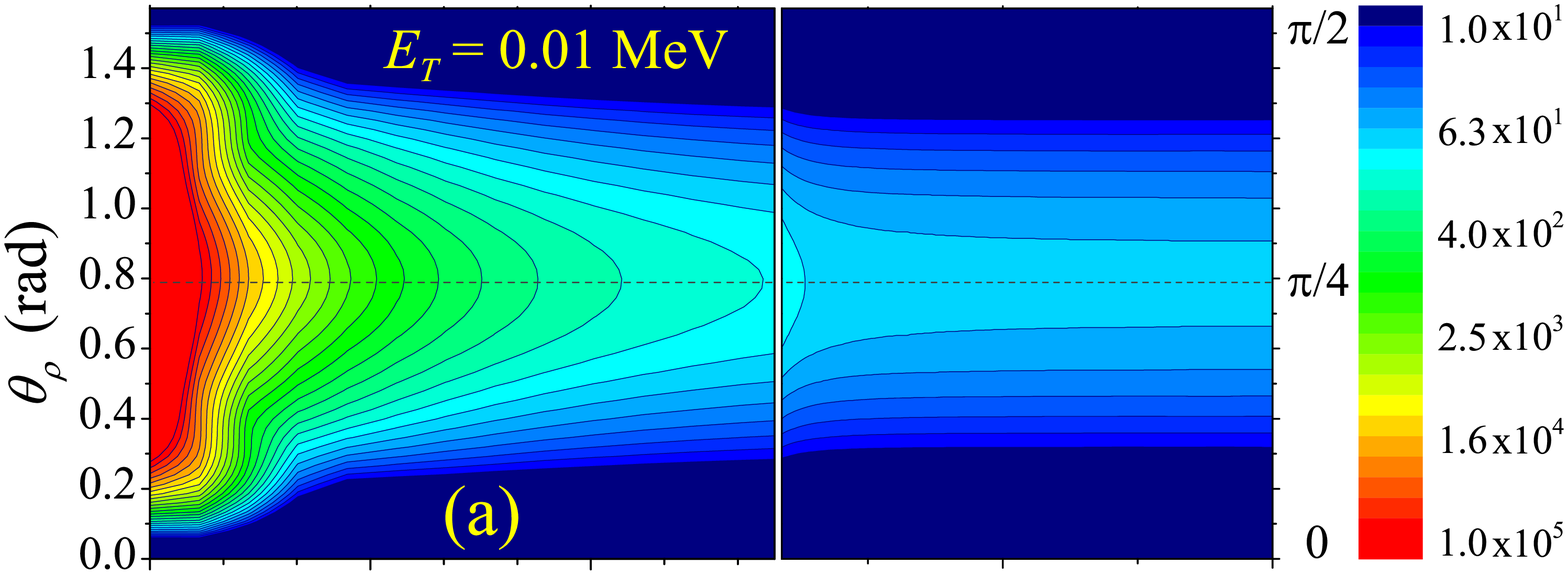}
\includegraphics[width=0.49\textwidth]{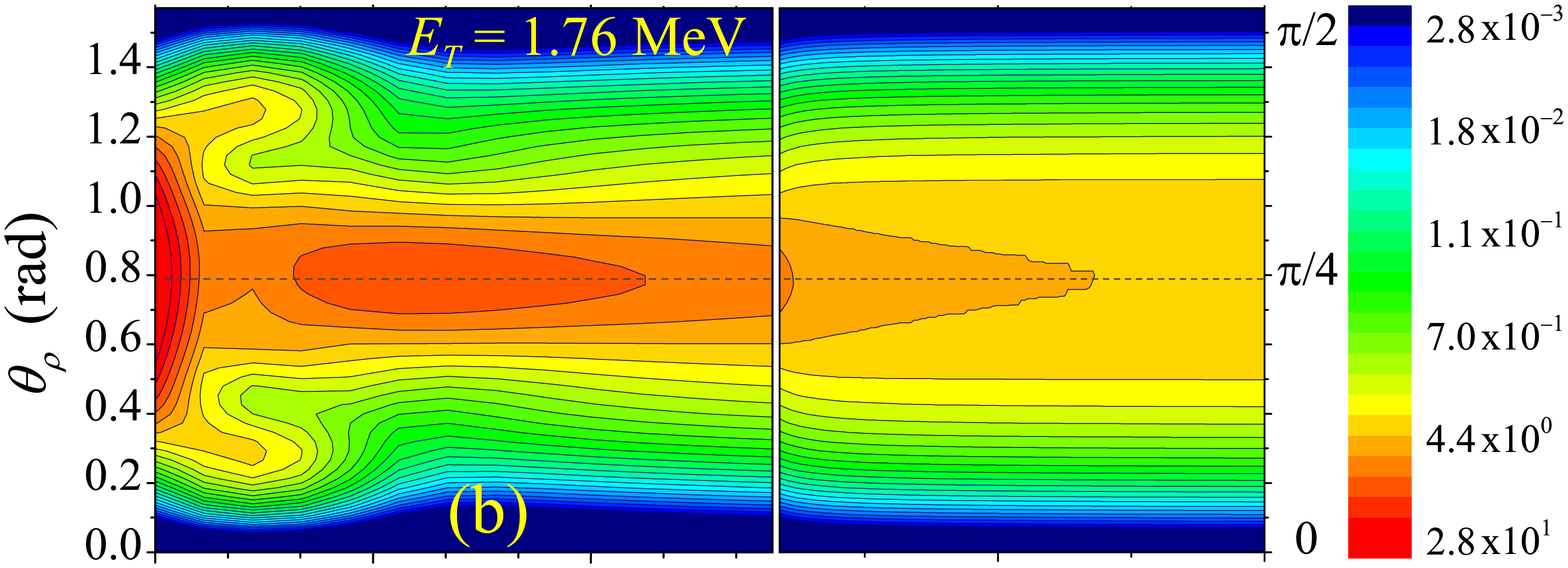}
\includegraphics[width=0.49\textwidth]{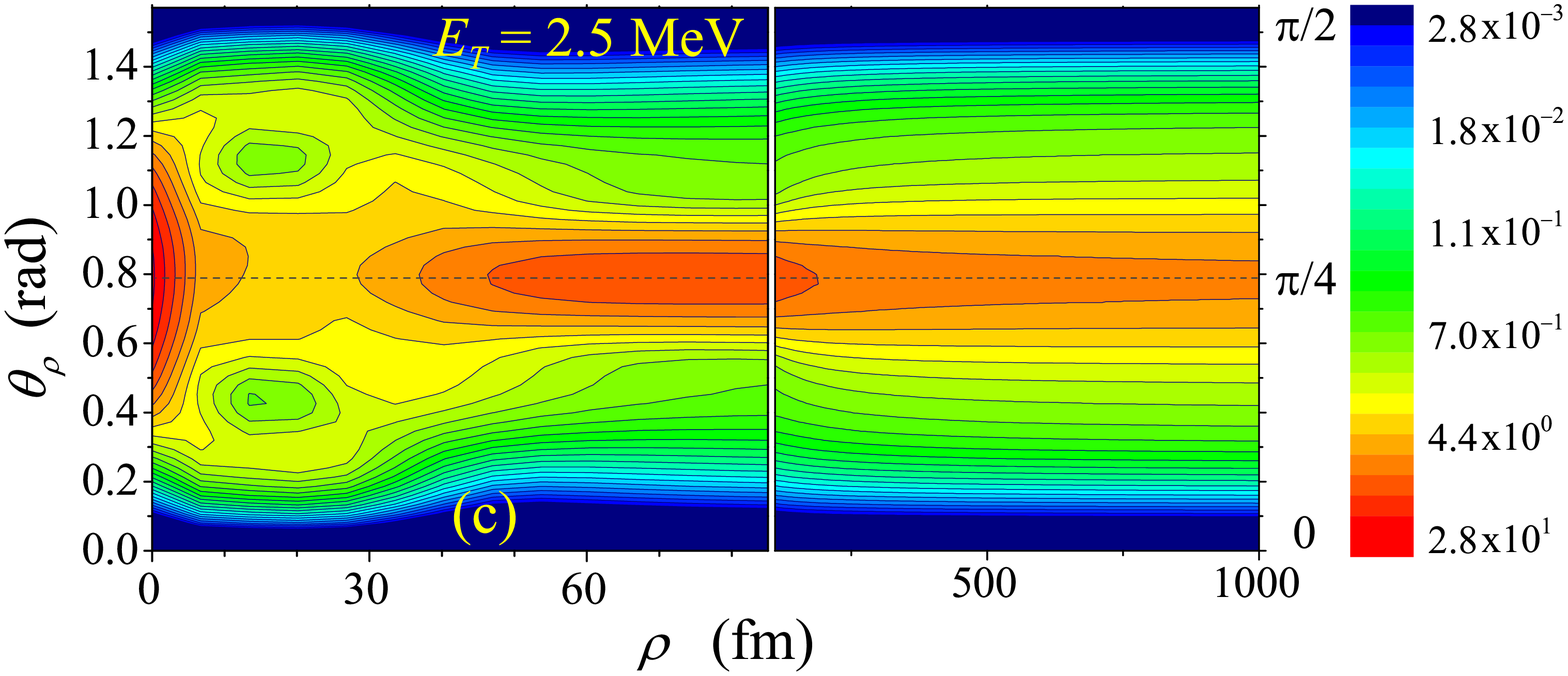}

\end{center}
\caption{(Color online) The $^{26}$O WF correlation density $W(\rho, 
\theta_{\rho})$ in the ``Y'' Jakobi system for the ground state (a) and two 
excited $0^+$ states (b) and (c). Notice the change of $\rho$ scale.}
\label{fig:wf-cor-den}
\end{figure}

To get a deeper insight into the process we may investigate the three-body WF
correlation densities Fig.\ \ref{fig:wf-cor-den}:
\[
W(\rho, \theta_{\rho}) = \int d \Omega_x \, d \Omega_y \, | \Psi^{(+)} (\rho,
\theta_{\rho},\Omega_x,\Omega_y) |^2 .
\]
In the ``Y'' Jacobi system we should choose the mass number of the $^{24}$O
cluster as $A_1$ or $A_2$ in Eq.\ (\ref{eq:coord}) and for such a relatively
heavy core cluster we can approximate the distances between core and valence
nucleons by:
\begin{eqnarray}
r_{\text{core-}n_1} &=& \sqrt{\textstyle \frac{A_1+A_2}{A_1A_2}} \, \rho \,
\sin(\theta_{\rho}) , \nonumber  \\
r_{\text{core-}n_2}   & \approx & \sqrt{\textstyle
\frac{A_1+A_2+A_3}{(A_1+A_2)A_3}} \, \rho \, \cos(\theta_{\rho}). \nonumber
\label{eq:rel-dis}
\end{eqnarray}
Thus, the provided correlation densities illustrate the evolution of relative
distances in the single-particle channels.

It can be seen in Fig.\ \ref{fig:wf-cor-den} that the correlation densities for
the calculated $0^+_2$ and $0^+_3$ states have a complicated correlation pattern
in the hyperangle $\theta_{\rho}$ at large distances $10< \rho < 30$ fm. Such a
triple-peak pattern should be connected with the important contribution of the
$[d^2]$ configuration in the above-the-barrier region. The $[d^2]$ configuration
is as expected dominant in the nuclear interior $\rho < 10$ fm, but for the
ground state it is suppressed under the barrier, see Fig.\ \ref{fig:wf-cor-den}
(a). In contrast, for the $0^+_2$ and $0^+_3$ states, illustrated in Figs.\
\ref{fig:wf-cor-den} (b) and (c), the $[d^2]$ component extends also to the
peripheral region forming the triple-peak patterns.

Figure \ref{fig:ls-evolution} demonstrates the evolution of the $0^+$
states with $V_{ls}$. This evolution provides some confirmation that the
``remnants'' of the broad monopole $0^+_4$  state is present in the spectrum
Fig.\ \ref{fig:spec-0p} at about 4 MeV. With $V_{ls}$ approaching zero,
the $0^+$ spectrum shifts downwards in energy providing a much narrower, and
thus well defined, $0^+_4$ state. For larger negative $V_{ls}$ values the
$0^+_4$ state increases in energy and is totally dissolved in the continuum.


\section{The $2^+$ states}
\label{sec:2p}

\begin{figure}
\begin{center}
\includegraphics[width=0.47\textwidth]{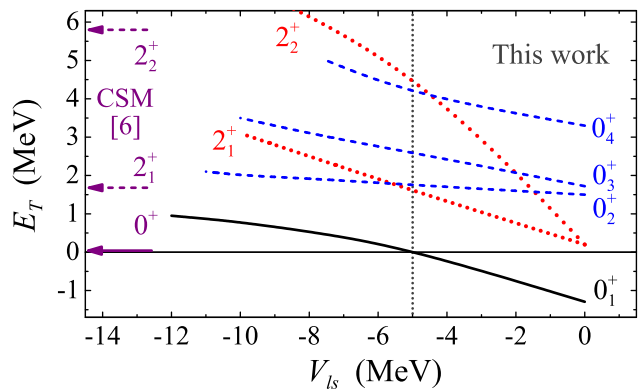}
\end{center}
\caption{(Color online) The systematics of the $0^+$ and $2^+$ states of 
$^{26}$O as a function of the $ls$ interaction intensity $V_{ls}$ in the 
$d$-wave $^{24}$O-$n$ channel under the condition that the $^{25}$O $d_{3/2}$ 
g.s.\ position remains fixed. The result for $V_{ls}=-5$ MeV (indicated by 
vertical dots) is provided in Fig.\ \ref{fig:levels} as our main result. The 
predictions of Ref.\ \cite{Volya:2006} are indicated by arrows in the left part 
of the plot.}
\label{fig:ls-evolution}
\end{figure}

The excitation spectrum of $^{26}$O is expected to have relatively low level
density due to the simplicity of the spectrum of $^{25}$O, where only one
low-lying state is known so far ($d_{3/2}$ at 0.7 MeV). In the calculations we
obtain just two $2^+$ states where the lower one has a structure characterized
by the configuration mixing $[d^2_{3/2}]_2$-$[d^2_{5/2}]_2$, while the higher
one is based on the $[s_{1/2}d_{3/2}]_2$-$[s_{1/2}d_{5/2}]_2$ configurations.
Within the three-body model the separation between these configurations is
defined by the $ls$ splitting between the $^{25}$O g.s.\ $d_{3/2}$ and the yet
unknown $d_{5/2}$ state. Figure \ref{fig:ls-evolution} demonstrates the
dependence of the positions of the $2^+$ states on the intensity of the $ls$
interaction under the condition that the $^{25}$O $d_{3/2}$ g.s.\ position
remains fixed. The $2^+$ states naturally become degenerate for $V_{ls} = 0$.

Most of the calculations of this work are performed for $V_{ls} =-5$ MeV,
which provides the position of the $0^+$ g.s.\ to be exactly on the $2n$
threshold. Under this condition the predicted positions of the $2^+$ state are
1.6 MeV and 4.5 MeV. The calculated three-body width value for the $2^+_1$ state
is
$\Gamma=115$ keV. However, the width of the states with expected dominant
sequential decay mechanism are not reliably predicted in the HH method
calculations. For that reason we also performed estimates using the direct decay
model expression Eq.\ (\ref{eq:sequent}). The result for the
$[s_{1/2}d_{3/2}]_2$ configuration which has the most favorable penetration
conditions is provided in Fig.\ \ref{fig:wid-estim}. For the $2^+_1$ state the
value $\Gamma_{1/2,3/2} \approx 2.7$ MeV. However, it should be corrected for
the weight of the $[s_{1/2}d_{3/2}]_2$ configuration in the interior of the
three-body WF $\Psi^{(+)}$,
\[
\Gamma = \Gamma_{j_1j_2} W_{j_1j_2} \,.
\]
The value $W_{1/2,3/2}=3.6\,\%$, which is quite small, is obtained in the
three-body calculations. Then the estimate $\Gamma=96$ keV is obtained. This
value agrees well with the results of the three-body width calculations, and we
estimate that it is realistic to expect the width of the first $2^+$ state to be
in the range $100-120$ keV.

\section{Discussion of theoretical results on $^{26}$O spectrum}


In our calculations we have chosen the calculation scheme, where the effect of
the occupied $d_{5/2}$ orbital in $^{25}$O is imitated by inverse $ls$ forces
moving the  $d_{5/2}$ state to an energy higher than that of the $d_{3/2}$
state. The particular value of this interaction was fixed by reproducing the
near-threshold position of the $^{26}$O g.s. Such a computation scheme is not
free of problems, but it seems, however, to be quite successful for the $^{26}$O
nucleus. This nucleus is just two neutrons away from the $^{28}$O neutron shell
closure and the main dynamical degrees of freedom are connected just with the
motion of the two valence nucleons. This is a clear motivation for the use of
the three-body core+$n$+$n$ model.

The oxygen isotope chain was studied in details by Volya and Zelevinskiy
\cite{Volya:2006} using the continuum shell model. They predicted the $^{26}$O
ground state position to be practically exactly on the threshold with $E_T=21$
keV. Such a bold prediction was nicely confirmed by the recent experimental
studies \cite{Lunderberg:2012,Caesar:2013,Kondo:2014}. The predictions of
\cite{Volya:2006} about $2^+$ states are in reasonable agreement with our
calculations: our $2^+_1$ state position is around 100 keV lower in excitation
energy than in Ref.\ \cite{Volya:2006} and, correspondingly, the $2^+_2$ is
about 1.7 MeV lower. The $0^+_{i>1}$ low-lying excited states are absent in the
continuum shell model calculations of Ref.\ \cite{Volya:2006}. This, probably,
could be connected to the ``short range character'' of the shell model
calculations in general: in our calculations the excited $0^+$ states require
the dynamical range of tens of Fermi to be properly accounted for. Another
possible reason could be that the treatment of $^{24}$O+$n$+$n$ continuum in
Ref.\ \cite{Volya:2006} is not fully dynamical, as it relies on the simplified
three-body Green's functions not including $N$-$N$ final-state interaction. For
this reason it is a astonishing that the width obtained in the current
calculations practically coincides with results the of \cite{Volya:2006}.
According to our experience, the simplified three-body Green's function should
provide a width for the $[s^2]$ configuration decay which is $\sim 10-30$ times
smaller than that obtained if appropriately taking the final state
nucleon-nucleon interaction into account.

The $^{26}$O g.s.\ was studied in \cite{Hagino:2014}. The higher $0^+$
excitations are not discussed in that work but there are some indications of the
excited $0^+$ at about 3.4 MeV, see, Fig.\ 1 of Ref.\ \cite{Hagino:2014}. The
authors do not elaborate this result and its reliability is not known; at least
there is some indication of the possibility of an existence of the low-lying
excited $0^+$ in an alternative approach. The first $2^+$ state was obtained at
about $E_T=1.35$  MeV in the recent calculations Ref.\ \cite{Hagino:2014b}. This
value is not drastically different from our results and the results of
\cite{Volya:2006}.

The ab-initio shell model theoretical calculations of Ref.\ \cite{Bogner:2014}
provided the excitation energy of the $2^+$ state in $^{26}$O in the range
$E^*=1.2-1.7$ MeV depending on the details of calculations. The obtained results
are reasonably consistent with other theoretical predictions.


\section{Discussion of relevance to experimental data}


How could the predicted excitations of $^{26}$O in our work be related to
observations? The considerable evolution of the $^{26}$O spectrum shape with the
$V_{ls}$ parameter is shown in Figure \ref{fig:ls-evolution}.
However, it should be noted that the predicted picture of \emph{excitation
energies} is relatively stable in this plot for reasonable variation of
$V_{ls}$. In particular, the excitation energies for $0^+_{i>1}$ states and
$2^+_1$ state are practically constant.

There is some evidence for intensity at about $E_T = 2$ MeV in Refs.\
\cite{Lunderberg:2012,Kohley:2013}. A peak just above 1 MeV was observed in
Ref.\ \cite{Kondo:2014}. For this energy range we predict relatively narrow
($\Gamma \sim 0.11$ MeV)  $2^+_1$ state at $E_T \sim 1.6 $  MeV. The right
``wing'' of this peak could be situated on a ``background'' formed by
$0^+_2$ and, maybe, $0^+_3$ states.

Evidence for the $^{26}$O excited state at about $E_T = 4.2$ MeV was obtained
in Ref.\ \cite{Caesar:2013}, although with marginal statistics. Within this
energy range we predict quite broad and overlapping $2^+_2$ and $0^+_4$ states.


\section{Conclusions}


We studied the $0^+$ and $2^+$ continuum properties of the $^{26}$O system in
three-cluster $^{24}$O+$n$+$n$  theoretical model. The main results obtained in
this work are:

\noindent (i) The ground state decay spectrum of the long-living $2n$ emitters
can be very well approximated in a broad energy range by a simple analytical
expression generalizing R-matrix phenomenology for use in the hyperspherical
space of three-body systems.

\noindent (ii) A number of monopole (breathing mode) excited $0^+$ states are
predicted in $^{26}$O at about 1.76, 2.6, and 4.2 MeV. These states extend to
extreme distances in radial space with typical single-particle orbital sizes
around $20$ fm. The predicted densely spaced sequence of $0+$ states in the
$^{26}$O continuum states with such radial properties resembles the Efimov
phenomenon. These states can probably be considered as the lowest Efimov states
forced into continuum by insufficient binding.

\noindent (iii) The predicted $2^+$ states in $^{26}$O at $1.6$ and $4.5$
MeV are in a reasonable agreement with continuum shell model calculations.

\noindent (iv) We suggest to identify the experimentally observed intensity in
the range $1-2$ MeV in the spectrum of $^{26}$O as a ``pileup'' of $0^+_2$ and
$2^+_1$ state contributions. The possible experimental peak at $\sim 4$ MeV in
$^{26}$O can be associated with overlapping broad $2^+_2$ and  $0^+_3$ continuum
states.

%
\textit{Acknowledgments.}
%
%
--- We are grateful to Prof.\ G.\ Nyman for careful reading of the manuscript
and numerous useful comments. L.V.G.\ is partly supported by Russian Foundation
for Basic Research Grant No.\ 14-02-00090-a and  Ministry of Education and
Science of the Russian Federation Grant No.\ NSh-932.2014.2.


\bibliographystyle{apsrev}
\bibliography{c:/latex/all}


\end{document}